\documentclass[aps,prc,superscriptaddress,showpacs,amssymb,amsmath,amsfonts,twocolumn,floatfix]{revtex4}
\usepackage{graphicx}

\begin{document}

\title{Properties of the $\Lambda$(1520) Resonance
  from High-Precision Electroproduction Data}
\newcommand*{\DUKE}{Department of Physics, Duke University
	    Triangle Universities Nuclear Laboratory, 
	    Durham, NC 27708, USA}
\newcommand*{\JLAB}{Thomas Jefferson National Acceleration Facility,
            Newport News, VA 23606, USA}
\newcommand*{\GWU}{Center for Nuclear Studies, Department of Physics, 
            The George Washington University, Washington, DC 20052, USA}
\newcommand*{\PNPI}{Petersburg Nuclear Physics Institute, Gatchina, 
            188300, Russia}
\newcommand*{\UVA}{University of Virginia, Charlottesville, VA 22901, USA}
\author {Y.~Qiang}
\affiliation{\DUKE}
\affiliation{\JLAB}
\author {Ya.I.~Azimov}
\affiliation{\PNPI}
\author {I.I.~Strakovsky}
\affiliation{\GWU}
\author {W.J.~Briscoe}
\affiliation{\GWU}
\author {H.~Gao}
\affiliation{\DUKE}
\author {D.W.~Higinbotham}
\affiliation{\JLAB}
\author {V.V.~Nelyubin}
\affiliation{\UVA}

\begin{abstract}
High-resolution spectrometer measurements of the reaction
$H(e,e'K^+)X$ at small $Q^2$ are used to extract the
mass and width of the $\Lambda(1520)$.  We investigate
the influence of various assumptions used in the extraction.
The width appears to be more sensitive to the assumptions
than the mass. To reach a width uncertainty about 1~MeV
or better, one needs to know the structure of the non-resonant
background.  Based on the new Jefferson Lab Hall A data,
our final values for
the Breit-Wigner parameters are $M=1520.4\pm0.6$(stat.)$\pm1.5
$(syst.)~MeV, $\,\Gamma=18.6\pm1.9$(stat.)$\pm1$(syst.)~MeV.
We also estimate, for the first time, the pole position for this
resonance and find that both the pole mass and width seem to be
smaller than their Breit-Wigner values.
\end{abstract}

\maketitle

\section{Introduction}
\label{sec:intro}

The $\Lambda(1520)$ is considered to be one of the best known
baryon resonances. It is, in particular, an excited baryon state
of the highest rank (4 stars) in the Review of Particle Physics
(RPP)~\cite{PDG}; but its properties are still not exceedingly
well understood.  For example, its decay properties~\cite{PDG},
with a rather large branching ratio to the $\pi\pi\Lambda$
channel, hint at an unexpectedly strong coupling to the kinematically
suppressed channel $\pi\Sigma(1385)$ that is stronger than to the
main decay channels $\pi\Sigma$ and $\overline{K}N$. This could
mean that the $\Lambda(1520)$ has
a molecular nature (see, \textit{e.g.}, Ref.~\cite{oset}
and references therein). On the other hand, evidence has
arisen~\cite{zou} that a new, previously unobserved, resonance
$\Sigma(1380)$ with $J^P=1/2^-$ and decay to $\Lambda\pi$ might
influence the mode $\Lambda(1520)\to\pi\pi\Lambda$. These examples
show that more detailed studies of the $\Lambda(1520)$, which may
clarify both its nature and the hyperon spectroscopy, are needed.

Surprisingly, the latest experimental inputs in RPP for the
mass and width of the $\Lambda(1520)$ are dated before 1980.
Since then, many new experiments
have been performed
and much more data
with the $\Lambda(1520)$ observed either as the main goal or
as a byproduct have been collected (even more data are expected
to appear in near future). For example, while
the main purpose of the recent Jefferson Lab (JLab)
Hall~A experiment~\cite{qi07}
was to search for $\Theta^+$-partners,
it also produced high quality data on the $\Lambda(1520)$.
The preliminary analysis~\cite{qi07} of this data
provided results for the mass and width of this
resonance that were in reasonable agreement with
the RPP values.

Note that most of the earlier high-precision mass and width values,
in particular those used for averaging in RPP~\cite{PDG}, had come
either directly from production measurements with bubble chambers,
or indirectly from partial wave analyses. High-energy spectrometers
did not have high enough resolution for such measurements, and
thus, their results could not compete with bubble chamber data,
even when collecting much higher statistics.  Now, the High
Resolution Spectrometers (HRS's)~\cite{HRS} in the Hall~A
provided an exciting accuracy ($\sigma=1.5$~MeV~\cite{qi07}),
comparable to the best previous resolutions. That is why we
reconsider here the Hall A data with
the aim to analyze the $\Lambda(1520)$ more accurately.

In this Letter, of course, we will not be able to answer all 
questions related to the $\Lambda(1520)$. We are going to
extract new values of its basic parameters, mass and width.
But, in addition to this standard procedure, we try also
to investigate how reliable may be those values, how strongly 
they can depend on various assumptions used. One more new 
point in this Letter is extraction of the pole position for 
the $\Lambda(1520)$.

\section{ JLab Hall~A Experiment}
\label{sec:expt}

High-resolution measurements of the missing mass (MM) spectra
were developed at near-forward production angles in the reactions
$ep\to e'K^\pm X$ and $e'\pi^+X$. The experiment~\cite{qi07} took
place in Hall~A at the Thomas Jefferson National Accelerator
Facility (JLab) using a 5.09~GeV electron beam incident on a 15~cm
liquid hydrogen target. Scattered electrons were detected in one
of the HRS's in coincidence with electroproduced
hadrons in the second HRS. Each spectrometer was positioned at
$12.5^\circ$ relative to the beamline, but the use of
additional septum magnets~\cite{septum} allowed to reach
smaller production angles, down to $\,\,\sim6^\circ$.  In this
configuration, the spectrometers have an effective acceptance of
approximately 4~msr in solid angle and $\pm$4.5\% in momentum,
while still maintaining their nominal $10^{-4}$
full-width half-max momentum resolution~\cite{HRS}.
To obtain the desired MM coverage, the central momentum of the
electron HRS was varied between 1.85 and 2.00~GeV, while the
central momentum of the hadron HRS was changed
between 1.89 and 2.10~GeV.  In such configuration, the average
momentum transfer of the virtual photon was \mbox{$<Q^2>~\approx
~0.1$~(GeV/c)$^2$}, and the average center-of-mass (CM) photon
energy was $<E^{\text{cm}}_{\gamma^*}>~=~1.1$~GeV which means
that $<$~W~$>$~=~2.53~GeV. For the kaon kinematics, the CM
scattering angle was
5.6$^{\circ}\le\theta^{\text{cm}}_{\gamma^*K}\le11.4^\circ$, and
the angular acceptance was
$\Delta\Omega^{\text{cm}}_{\gamma^*K}\approx 38$~msr.

Calibration of HRS's was based on precise measurements of the
known MM peaks for the neutron (in $\pi^+X$) and for the hyperons
$\Lambda$(1116), $\Sigma^0(1193)$ (both in $K^+X$). These three
baryons decay through weak interactions, and the proper widths of
their peaks are negligible. Therefore, the observed widths of the 
peaks directly determine the MM resolutions at the corresponding
momenta of the registered meson ($\pi^+$ or $K^+$). Then, the
resolution of HRS for the $\Lambda(1520)$ may be determined by
extrapolation of those resolution values to the MM region of 
1520~MeV. Such a procedure gave the resolution $\sigma=
1.5~\textrm{MeV}$~\cite{qi07}. We apply this resolution 
when fitting the data.

\begin{figure}[th]
\centerline{
\includegraphics[height=0.3\textwidth, angle=90]{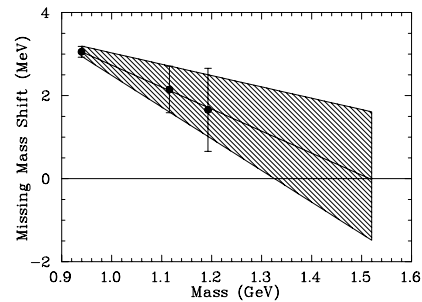}}
\vspace{3mm}
\caption{Mass calibration using the missing mass measurements
         of the neutron, $\Lambda(1116)$, and $\Sigma^0(1193)$.
         The vertical axis shows deviation of the measured 
         mass from the RPP mass value~\protect\cite{PDG} 
         given at the horizontal axis. The linear best-fit 
         result (the central solid line) is extrapolated 
         to the $\Lambda$(1520) position. The hatched area
         characterizes possible errors of the extrapolated 
         correction to the scale calibration. 
         \label{fig:g1}}
\end{figure}

Positions of the peaks, as measured by the HRS, show deviations 
from the RPP mass values, different for different peaks (see 
Fig.~\ref{fig:g1}). We use those shifts to correct the 
spectrometer MM scale at the corresponding mass value. Then, 
mass correction for the peak of $\Lambda(1520)$ may be obtained 
by extrapolation of the peak shifts to the MM region of 1520~MeV. 
As seen in  Fig.~\ref{fig:g1}, the measured shift values admit 
simple linear extrapolation providing negligible correction
near 1520~MeV. 

Let us estimate uncertainty of this zero correction for the mass 
scale near $\Lambda(1520)$. The same peak positions, being 
measured in 4 different parts of the HRS, show some spread of 
the mass values. It contributes, of course, to systematic 
uncertainty of any mass measurement by the HRS, but appears 
to be small, less than 0.2~MeV, \textit{i.e.}, smaller than 
the statistical errors shown by error bars in Fig.~\ref{fig:g1}. 

More essential for the mass scale beyond location of the three 
peaks is another source of uncertainty, related to the 
extrapolation procedure. Statistical errors of the particular 
shifts (mainly of hyperon ones) in Fig.~\ref{fig:g1} spreads 
the possible slope for the linear extrapolation. As a result, 
the MM scale correction near 1520~MeV may be non-zero, having 
uncertainty about $\pm1.5$~MeV. We will consider this value 
as the possible systematic uncertainty for the position of 
the $\Lambda(1520)$. Further experimental details can be 
found in Refs.~\cite{qi07,qi07T}.

\section{Current parameters of the $\Lambda(1520)$ }
\label{sec:curr}

Before we proceed further with our analysis, let us briefly
consider earlier results (Table~\ref{tab:tbl1}) that were used
in the RPP averaging~\cite{PDG}. The different values for the 
mass and width of the $\Lambda(1520)$ look to be in reasonable 
agreement with each other and with the current average.

\begin{table}[th]
\caption{$\Lambda(1520)D_{03}$ parameters.
         \label{tab:tbl1}}
\vspace{2mm}
\begin{tabular}{|c|c|c|c|c|}
\hline
Mass           & Width        & Evts  &Year & Ref. \\
(MeV)          & (MeV)        &       &     &   \\
\hline
1517.3$\pm$1.5 & 16.3$\pm$3.3 &  300  &1980 & \protect\cite{ba80} \\
1517.8$\pm$1.2 & 14  $\pm$3   &5k/677 &1979 & \protect\cite{ba79} \\
1519.7$\pm$0.3 & 16.3$\pm$0.5 &  4k   &1977 & \protect\cite{ca77} \\
1519.4$\pm$0.3 & 15.5$\pm$1.6 &  2k   &1975 & \protect\cite{co75} \\
1520.0$\pm$0.5 & 15.4$\pm$0.5 & PWA   &1978 & \protect\cite{al78} \\
1519  $\pm$1   & 15.0$\pm$0.5 & PWA   &1977 & \protect\cite{go77} \\
\hline \hline
avr. 1519.5$\pm$1.0 & avr. 15.6$\pm$1.0 & & 2008 & \protect\cite{PDG} \\  \hline
\end{tabular}
\end{table}

The corresponding works fit the data by accounting for both the
resonance itself and a background. However, they use different
approaches to this task. Most papers deal with the resonance
production~\cite{ba80,ba79,ca77,co75}. They study various
inelastic reactions, with various many-particle final states, and
look for a peak in the mass distribution of one or another final
subsystem. The distribution is described by a Breit-Wigner (BW)
contribution with some non-coherent (non-interfering) background.
Any possible interference with other resonances is neglected as
well.  Now we consider in some detail the difference in treatments 
of various production measurements.

Paper~\cite{ba80} considers only the $K^-p$ decay channel, with 
the relativistic BW term having the momentum-dependent width and 
with the background linear in $M(K^-p)$. On the other hand, the 
authors of Ref.~\cite{ba79} obtain the $\Lambda(1520)$ mass mainly 
from mass distributions in different decay channels, $\pi^-\Sigma^+$ 
and $\pi^+\Sigma^-$, with small addition of $\Lambda\pi^+\pi^-$ 
events. The resonance width is extracted only from the 
$\Lambda\pi^+\pi^-$ events. Here, the BW term is presented in the 
non-relativistic form, with the energy-independent width; the 
background is taken to be linear in M($\pi\Sigma$) or 
M($\Lambda\pi^+\pi^-$).

Paper~\cite{ca77} studies several final subsystems, but
extracts the best (and final) values for the $\Lambda(1520)$ mass
and width from the $K^-p$ mass spectra. It uses the relativistic
form for the BW term, with the energy-independent width. The
background is assumed to be linear in $M^2(K^-p)\,$.

Paper~\cite{co75} mainly studies also the decay channel
$K^-p\,$, the width is assumed to be energy-dependent,
accounting for the appropriate angular-momentum barrier. More
details of the BW term and, especially, background are not
described explicitly.

On the other hand, papers~\cite{al78, go77} study the direct
resonance formation. They construct partial wave analyses for the
elastic $K^- p\,$ scattering, with some way (different in the two
papers) of accounting for inelastic processes. Separation of
resonance/background is made in a particular partial amplitude, so
their contributions are coherent. Both papers use the
non-relativistic BW form and include the angular momentum barrier
into the width, though in different ways. The background is also
described differently in the two papers.

Without any justification, all those papers, which assume the
energy dependent width of the $\Lambda(1520)$, describe the
total width as having the threshold behavior related only to
the $K^-p\,$ channel. Meanwhile, different thresholds,
corresponding to other decay channels (\textit{e.g.},
$\pi\Sigma$), should also affect the total width.

All the described approaches would be completely equivalent for
an ideally narrow resonance. But, for the case of a finite width,
such as the $\Lambda(1520)$, they should provide different 
results at some level of accuracy. In what follows, we not only 
extract the resonance parameters from the Hall A measurements, 
but also estimate the sensitivity of our results to how the 
experimental data was treated.

\section{Present analysis of the $\Lambda(1520)$ }
\label{sec:analys}

The earlier Hall~A paper~\cite{qi07} analyzed the inclusive 
reactions
\begin{equation}
	\gamma^* + p\to K^+(\pi^+) + X   \label{reac}
\end{equation}
mainly to search for narrow (presumably, exotic) peaks in the 
missing mass (\textit{i.e.}, $M_X$), with widths $\,\sim1$~MeV 
or less. Respectively, the experimental data were presented 
with the narrow binning in $M_X\,$: the width of one bin was 
1~MeV. For the case of $K^+$, the MM distribution showed, of
course, the pronounced peak of the known resonance 
$\Lambda(1520)$. It was analyzed, but only in a preliminary
way, to demonstrate reasonable agreement of its mass and 
width with their RPP values~\cite{PDG} and, thus, to confirm
satisfactory understanding of the mass scale and resolution.
Here we reanalyze the $\Lambda(1520)$ more accurately.

\begin{figure}[th]
\centerline{
\includegraphics[height=0.6\textwidth, angle=0]{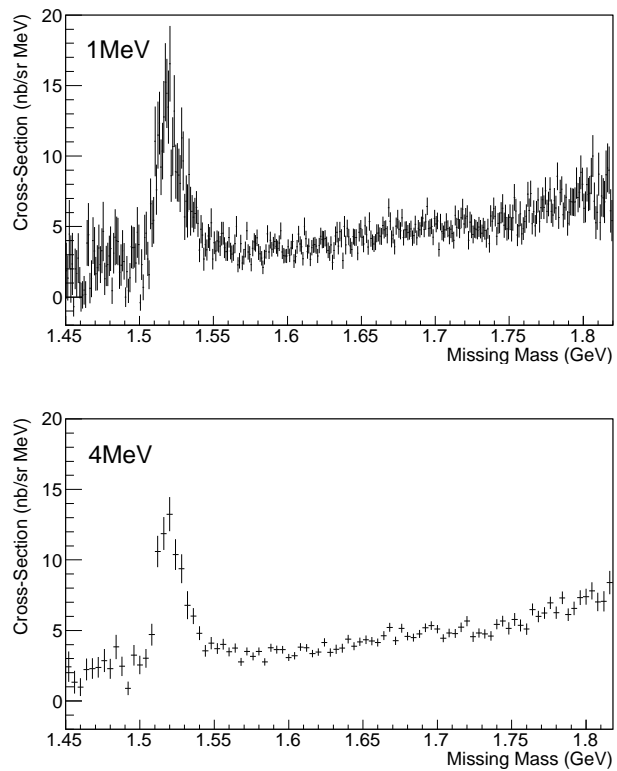}}
\vspace{3mm}
\caption{Missing mass distribution in the reaction $\gamma^*p\to
        K^+X$~\protect\cite{qi07} with two different binnings: 
	1~MeV in the upper panel; 4~MeV in the lower panel. 
	\label{fig:g2}}
\end{figure}

First of all, we modify data presentation. The 1~MeV steps 
for $M_X$, used in Ref.~\cite{qi07}, were appropriate to scan 
in searching for resonances of $\sim1$~MeV width. But bins 
of such a size are unsuitable to study the resonance of 
$\sim16$~MeV width with Gaussian resolution 1.5~MeV. Too 
narrow bins provided stronger fluctuations and excessively 
large statistical errors 
for every experimental point. That is why in the present 
reanalysis of the data for $\Lambda(1520)$ we increase the 
bin size from 1~MeV to 4~MeV. This size is more adequate for 
the resolution with $\sigma=1.5$~MeV, it essentially 
diminishes statistical errors and fluctuations. This is seen 
in Fig.~\ref{fig:g2}, which displays the cross section 
$d\sigma/(dM_X\,d\Omega_{\gamma^*K})$ as a function of $M_X$ 
for two different bin sizes. In what follows, we use the 
distribution in the lower panel of Fig.~\ref{fig:g2} to 
extract the $\Lambda(1520)$ mass and width.

The system $X$ in the shown mass region may consist of either 
$N\overline{K}$, or $\Sigma\,(\Lambda)$ hyperon, accompanied 
by pion(s). Any resonance is revealed in the MM distribution 
inclusively, being summed over all possible final states.
The width of $\Lambda(1520)$ also consists of several
contributions from different decay modes. Their energy 
dependence should be different, which was not accounted for 
in the preliminary analysis~\cite{qi07}, but will be taken 
into account now. Furthermore, in contrast to all earlier 
works, we will try various manners of data treatment and 
consider their influence on the resulting extracted values 
of the resonance parameters. 

Generally, we describe the MM spectra in the form
\begin{equation}
	{\rm Fit} = BW + BG \,, \label{fit}
\end{equation}
where $BW$ is the Breit-Wigner contribution for the resonance
$\Lambda(1520)$, and the term $BG$ combines all other contributions
(including other possible resonances), which provide a
background for the $\Lambda(1520)$.

The BW contribution may be written as
\begin{equation}
	BW = A_{BW}~\Gamma(M_X)~D(M_X)\,. \label{res}
\end{equation}
In the non-relativistic form, we have
\begin{equation}
	D^{-1}_{nrel}(M_X)=(M_X - M_0)^2
       + \Gamma^2(M_X)/4\, .
\label{nrel}
\end{equation}
In the relativistic form, this denominator should
look like $|M_X^2 -[M_0 - i\Gamma(M_X)/2]^2\,|^2 $,
but usually  $\Gamma^2\ll M^2_0$, and $D^{-1}$
may be approximately written in the form
\begin{equation}
	D^{-1}_{rel}(M_X)= (M_X^2-M_0^2)^2 + M_0^2\, \Gamma^2(M_X)\,.
\label{rel}
\end{equation}
With the energy-independent width, we define $\Gamma(M_X)\equiv
\Gamma_0\,$. Here $M_0$ and $\Gamma_0$ may be considered as the BW
mass and width of the $\Lambda(1520)$. It is definitely so for the
non-relativistic form; more detailed discussion of the BW parameters
for the relativistic form will be presented in section~\ref{results}.

The energy-dependent width has a more complicated structure.
Let us recall that the total width is the sum of partial ones for
all decay channels, $\Gamma=\sum_i \Gamma_i\,$. Here we emphasize
that every partial width should have its own energy dependence,
corresponding to the threshold and kinematical properties of the
particular decay channel.

For a two-particle decay of the initial state of mass $M_X$, the
threshold behavior is proportional to $p^{1+2l}$, where $p$ is the
CM momentum of the decay products, $l$ is their orbital angular
momentum. The channels $\overline{K}N$ and $\pi\Sigma\,$ in decays
of the resonance $\Lambda(1520)$ with $J^P=3/2^-$ have $l=2\,.$
Therefore, for their partial widths we can write
\begin{equation}
	\Gamma_{\overline{K}N}(M_X)=\Gamma_0~B_{\overline{K}N}
	\left(\frac{p_K}{p_{0K}}\right)^5\left(\frac{M_0^2}
	{M_X^2}\right)^2 \,,
\label{N}
\end{equation}
\begin{equation}
	\Gamma_{\pi\Sigma}(M_X)=\Gamma_0~B_{\pi\Sigma}
	\left(\frac{p_\pi}{p_{0\pi}}\right)^5\left(\frac{M_0^2}
	{M_X^2}\right)^2\,,
\label{S}
\end{equation}
where $B_i$ are the branching ratios for the corresponding decay
channels, $p_K$ and $p_\pi$ are the corresponding momenta in the
rest-frame of the mass $M_X$, while $p_0$ are those momenta at
$M_X=M_0\,$. The form of the last factors is chosen so to not
violate the analyticity of amplitudes in the nearby $M_X^2$ region
and to have a non-increasing ratio $\Gamma(M_X)/M_X$ even at
$M_X\to\infty$.

Three-particle final states are described by several parameters,
so they provide more complicated structure to the corresponding
partial widths. But the threshold behavior is rather simple,
proportional to $(\sqrt{T})^{4+2L}$, where $T$ is the released
kinetic energy, $L$ is the sum of final orbital momenta (the first
term in the exponent is due to the 3-particle phase space). To
have $J^P=3/2^-$, the system $\pi\pi\Lambda$ needs at least one
$P$-wave, so $4+2L=6$, and we may write
\begin{equation}
	\Gamma_{\pi\pi\Lambda}(M_X)=\Gamma_0~B_{\pi\pi\Lambda}
	\left(\frac{T(M_X)}{T(M_0)}\right)^3\left(\frac{M_0^2}
	{M_X^2}\right)\,, \label{L}
\end{equation}
with $T(M_X)=M_X-M_\Lambda-2M_\pi\,$.

Decays of the $\Lambda(1520)$ are dominated by only three channels,
$\overline{K}N,\,\pi\Sigma,$ and $\pi\pi\Lambda\,$. Branching
ratios for two other channels, $\pi\pi\Sigma$ and $\Lambda\gamma$,
are not more than 1\% each~\cite{PDG}, and we neglect them here.
Also neglected is the channel $\Sigma^0\gamma$, with an even lower
branching ratio~\cite{PDG}. For the main channels we use the
branchings
\begin{equation}
	B_{\overline{K}N}=0.46\,,~~~B_{\pi\Sigma}=0.43\,,~~~
	B_{\pi\pi\Lambda}=0.11\,,
\label{Br}
\end{equation}
which agree with the RPP tables~\cite{PDG}. Summing over all decay
channels, we have $\Gamma(M_0)=\Gamma_0$.

Now we need to describe the background. Here we also discuss
various possibilities. First of all, background $BG$ consists
of several terms.

Contribution of the $S$-wave $\overline{K}N$ continuum
to the MM spectrum is described by the term
\begin{equation}
	BG_N=A_{\overline{K}N}\sqrt{M_X-M_{th}}\,,~~~M_{th}=M_N+M_K\,,
\label{BgN}
\end{equation}
while $D$-wave of this system is contained in the BW term of
the $\Lambda(1520)$ with the energy-dependent width. We neglect
the $P$- and higher-wave contributions.

The essential hyperon-meson contributions have lower thresholds,
and their non-resonance contribution near the $\Lambda(1520)$
may be approximated by a combination of three simple terms:
\begin{equation}
	BG_{Y0}= A_{Y0}=\textrm{const}\,, \label{BgY0}
\end{equation}
\begin{equation}
	BG_{Y1}=A_{Y1}\,(M_X-M_{th})\,,
\label{BgY1}
\end{equation}
\begin{equation}
	BG_{Y2}=A_{Y2}\,(M_X-M_{th})^2\,.
\label{BgY2}
\end{equation}
In addition to such smooth contributions, MM spectra in
the area of $M_X\approx1520$~MeV may also contain
contributions of additional resonances, except the presently
studied $\Lambda(1520)$. First of all, we mean the
$\Lambda(1405)$, which decays nearly totally into $\Sigma
\pi$~\cite{PDG}. But in our inclusive measurements with
$M_X>1.45$~GeV, it cannot be separated from the
$\Sigma(1385)$, with the dominant decay into $\Lambda\pi$ and
the 10\% level for $\Sigma\pi$~\cite{PDG}. Moreover, the
$N\overline{K}$ channel may also contain contributions of
the two resonances, as sub-threshold ones (through their
BW tails). Our data do not separate decay channels, and,
after all, we write background from such resonance contributions
in the form
\begin{equation}
	BG_R = A_R ~\Gamma_R~D_R(M_X)\,,
\label{BGR}
\end{equation}
with
\begin{equation}
	D^{-1}_R=
	(M_X^2-M_R^2)^2 + (M_R\, \Gamma_R)^2\,.
\label{R}
\end{equation}
Here we neglect energy dependence of the width and use
parameters of the $\Lambda(1405)$:
$$M_R=1406~\textrm{MeV}\,,~~~\Gamma_R=50~\textrm{MeV}\,,$$
in accordance with the RPP tables~\cite{PDG}.

Above the $\Lambda(1520)$, the tables of RPP~\cite{PDG} also
present several resonances, both $\Lambda$- and $\Sigma$-like
(with different star status). But our MM spectra (Fig.~\ref{fig:g2})
do not show any reliable manifestations of resonances, at
least up to $\,\sim1670$~MeV, and we will not take them into account
explicitly.

Thus, for extracting the mass and width of the $\Lambda(1520)$, we
use the MM spectrum shown in the lower panel of Fig.~\ref{fig:g2}
and restrict ourselves to the mass interval from 1.45~GeV to
1.65~GeV (this corresponds to 13,070 detected events). To the
peak, we apply the BW term (\ref{res}), while the background is
described by various combinations of terms
(\ref{BgN})-(\ref{BGR}). In all cases, we use the normalizing
coefficients $A_i$ as free fitting parameters, together with the
BW parameters $M_0$ and $\Gamma_0$. Further, for the best-fit
procedure, we use two not quite equivalent methods, least-squares
(min-$\chi^2$) and log-likelihood (LL) ones. In such a way, we
obtain several sets of numerical values for pairs
($M_0,\,\Gamma_0$) and can trace their dependence on the
assumptions used.

\section{Results and their Discussions}
\label{results}

We begin with considering changes of the BW mass $M_0$. They
appear to be rather small, though not always negligible. The
non-relativistic expression (\ref{nrel}) for the BW term provides
$M_0$ values lower than the relativistic Eq.(\ref{rel}), at the
level of $\,\sim0.05$~MeV.

Similarly, the LL procedure gives lower $M_0$ than min-$\chi^2$;
the difference may be as small as 0.01~MeV, but may reach
$\,\sim0.15$~MeV. Structure of the background can also shift
$M_0\,$; the difference for various variants which we use is,
again, not more than 0.15~MeV.

A more essential effect comes from energy dependence of the width.
Description with the energy-dependent width results in $M_0$ about
0.6~MeV lower than for the energy-independent width. Interestingly,
it is at the same level as our statistical uncertainty for $M_0$,
which is \mbox{$\,\sim0.6$~MeV} in all the studied cases.

Changes of the BW width $\Gamma_0$ also show some regularities.
Shifts of results for using the (non-)relativistic Eqs.(\ref{nrel})
or (\ref{rel}) are not more than 0.06~MeV. The LL fitting gives a
lower width than min-$\chi^2$; the difference may be 0.2~MeV, but
may reach $\,\sim0.9$~MeV (for comparison, our statistical
uncertainty for the width is $\,\sim2$~MeV).

More complicated is the influence of the background description.
By changing different possible combinations of background terms
(\ref{BgN})-(\ref{BgY2}), we change $\Gamma_0$, but not very
strongly, not more than 0.2~MeV. More influential is the resonance
$\Lambda(1405)$. Its exclusion diminishes $\Gamma_0$; the
difference may be up to $\,\sim1$~MeV (the corresponding shift of
$M_0$ is much smaller, not more than 0.1~MeV).

Now we are able to formulate some conclusions, which may have more
general character.
\begin{itemize}

\item{The width of $\Lambda(1520)$ is sufficiently small, so the
relativistic and non-relativistic forms give practically the same
values of $M_0$ and $\Gamma_0$ (at the present level of accuracy).}

\item{By definition, the LL fitting always provides a larger value
of $\chi^2\,$, than the min-$\chi^2$ fitting. However, formally
they should be equivalent at asymptotically high statistics. In
this sense, the present statistics is not asymptotical yet. Both
$M_0$ and $\Gamma_0$ are different in the two fittings, the
differences are comparable to the statistical uncertainties of
those BW parameters. In terms of $\chi^2$ per degree of freedom,
$\chi^2$/dof, which is typically $\,\sim1.5$ in our studies here,
the LL fitting is up to 0.05 higher than min-$\chi^2\,$.}

\item{The $M_0$ has a smaller statistical uncertainty and is less
affected by any change of fitting procedure than the $\Gamma_0$.}
\end{itemize}

The last point may be understood as follows. The spectrum in
the lower panel of Fig.~\ref{fig:g2} shows the clear peak for
$\Lambda(1520)$, and the value of $M_0$ is mainly determined by
the position of the peak maximum, with rather small statistical
uncertainty $\Delta M_0$. Therefore, any change of the fitting
procedure may provide a shift of the $M_0$ not larger than the
$\Delta M_0$.

This is not quite similar for the case of width. If neglecting
energy dependence, $\Gamma_0$ is the full width of the peak at its
half-height. In terms more directly related to the inclusive MM
spectrum, the peak half-height can be formulated as the half-depth
from the peak maximum. Thus, though the flanks of the peak are
rather well-determined, to extract $\Gamma_0$ one needs to know
the necessary depth, or the proper height of the peak, for the
given experimental spectrum. Of course, it essentially depends on
the height of the background under the peak. As a result, both
statistical uncertainty $\Delta\Gamma_0$ and changes of $\Gamma_0$
with fitting changes should be larger than those of $M_0$.
Moreover, we can admit that changes of $\Gamma_0$, with changes of
fitting procedure, might exceed the $\Delta\Gamma_0$ (though this
is not the case in our present study).

It is interesting that a change of the mass interval used for
fitting may also change the height of the background and, hence,
the $\Gamma_0$. If we take the smaller interval 1.45--1.60~GeV for
our fits, the value of $M_0$ may shift, but in all cases less than
0.05~MeV, while the value of $\Gamma_0$ diminishes with a stronger
shift, up to $\,\sim 0.5$~MeV.

\begin{figure}[th]
\centerline{
\includegraphics[height=0.4\textwidth, angle=0]{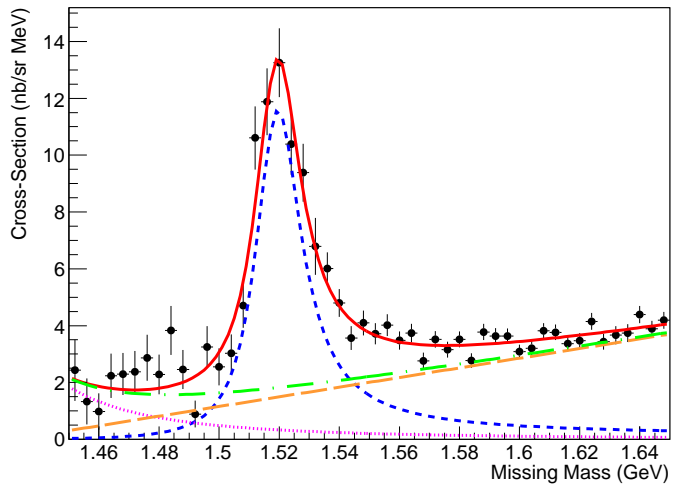}}
\vspace{3mm} \caption{(Color on line) Fit to the experimental MM 
	distribution (the lower panel of Fig.~(\ref{fig:g2})) is 
	shown by the solid line. The short-dashed line is the BW 
	contribution for the $\Lambda(1520)$ (note an asymmetric 
	form, due to the energy-dependent width). The total 
	background (dot-dashed line) is the sum of the linear 
	binomial long-dashed line) and tail of the $\Lambda(1405)$ 
	(dotted line). \label{fig:g3}}
\end{figure}

For extracting our resulting BW parameters, we use the BW
relativistic expression (\ref{res}),\,(\ref{rel}) with
energy-dependent width (both points are theoretically
motivated to be more reasonable). As a simple, but typical
representation for the background,  we combine the terms
(\ref{BgY0}),\,(\ref{BgY1}) and (\ref{BGR}),\,(\ref{R}). The
corresponding least-squares (min-$\chi^2$) fit in the mass
interval 1.45 -- 1.65~GeV is shown in Fig.~\ref{fig:g3}, together
with all separate contributions. This fit corresponds to
$\chi^2$/dof=1.46 and gives
\begin{equation}
	M_0 = 1520.4\pm 0.6~{\rm MeV}\,,~~ \Gamma_0 = 18.6\pm 1.9~{\rm
	MeV} \,, \label{MG}
\end{equation}
with pure statistical uncertainties. Each of the two BW
parameters has also a systematic uncertainty. It is about 
1.5~MeV, mainly due to the mass scale uncertainty, for $M_0$, 
and about 1~MeV, mainly due to the fitting procedure
ambiguity, for $\Gamma_0$.

For comparison, the same description with the LL fit gives
slightly lower values
\begin{equation}
	M_0 = 1520.3\pm 0.6~{\rm MeV}\,,~~ \Gamma_0 = 17.8\pm 1.9~{\rm
	MeV} \,, \label{MGL}
\end{equation}
and the higher value $\chi^2$/dof=1.50 . This LL fit agrees 
with the values $M_0=1519.9\pm0.6$~MeV and $\Gamma_0=16.5\pm
1.7$~MeV of the preliminary analysis~\cite{qi07} which also used 
the LL fitting (its systematic uncertainty was estimated to be
3~MeV for the mass; no estimations for the width). The 
difference between the present and preliminary LL fits is
essentially related to excessive fluctuations and statistical 
errors of the experimental points with 1~MeV binning.

Even without accounting for the systematic uncertainty, both $M_0$
and $\Gamma_0$ are in reasonable agreement with previous
experimental values and with their RPP average (see
Table~\ref{tab:tbl1}). It is worth to note, however, that the
uncertainties in the later works~\cite{ba80} and~\cite{ba79} are
larger than those in all the earlier works. This may hint that the
uncertainties stated in the earlier works (and, therefore, in the
average) are too optimistic.

In respect to experimental points, the fit in Fig.~\ref{fig:g3}
looks reasonably good. One can note, however, some small but
regular excess of the experimental values over the fit near
$M_X=1.48$~GeV. It could be due to a small contribution of
$\Sigma(1480)$, not accounted for in our fit. This contradictory
state, with low (one-star) status in the RPP~\cite{PDG}, recently
obtained new experimental support~\cite{anke}. Manifestations of
the $\Sigma(1480)$ seem to be seen also in several other recent
papers~\cite{herm,svd1,zeus,svd2}. We have tried to include this
resonance into our fit, coherently or incoherently, but large
uncertainties prevent us from any conclusions.

Now we can discuss position of the $S$-matrix pole corresponding
to the $\Lambda(1520)$, that has never been discussed for hyperons. 
It is determined by vanishing of the denominator (\ref{rel}), 
\textit{i.e.}, by a solution of the equation
\begin{equation}
	(W_p^2-M_0^2)^2 + M_0^2\, \Gamma^2(W_p)=0\,. \label{pole}
\end{equation}
This non-linear equation may have non-unique solutions. Physically
reasonable is only one of them, close to $(M_0-i\, \Gamma_0/2)\,$.
It is convenient, therefore, to rewrite Eq.(\ref{pole}) in the
form
\begin{equation}
	W_p =M_0 \left[1- i\,\Gamma(W_p)/M_0\,\right]^{1/2}\,. \label{pol}
\end{equation}
Its complex solution gives the pole mass $M_p= \textrm{Re}\,W_p$
and the pole width $\Gamma_p=-2\,\textrm{Im}\,W_p$.

For the energy-independent case $\Gamma(W)\equiv\Gamma_0\,$, the
arising $W_p$ value (\ref{pol}) provides exact BW parameters. Up
to corrections of the relative order $(\Gamma_0/M_0)^2$, more
exact values for the BW mass and width are
\begin{equation}
	M_{BW}=M_0[\,1+\frac18\,(\Gamma_0/M_0)^2\,]\,; \label{BWm}
\end{equation}
\begin{equation}
	\Gamma_{BW}=\Gamma_0[\,1-\frac18\,(\Gamma_0/M_0)^2\,]\,. \label{BWg}
\end{equation}
The differences between $M_{BW}$ and $M_0\,$, $\Gamma_{BW}$ and
$\Gamma_0$ arise due to the approximate character of expression
(\ref{rel}). The corresponding shifts are numerically small, much
smaller than the statistical uncertainties (\ref{MG}).

If we assume the width to be energy-dependent and take it as the
sum of expressions (\ref{N}), (\ref{S}), and (\ref{L}), then, by 
definition, $\Gamma(M_0)=\Gamma_0$. Of course, the solution of 
Eq.(\ref{pol}) still corresponds to the pole position, but it 
differs from the simple BW position of the energy-independent case. 
Numerical solution with values (\ref{MG}) for $M_0$ and $\Gamma_0$ 
gives
\begin{equation}
	M_p=1518.8~\textrm{MeV} \,,~~~ \Gamma_p=17.2~\textrm{MeV}\,.
\label{MpGp}
\end{equation}
Note that $M_p<M_{BW}$, $\Gamma_p<\Gamma_{BW}$, with the mass
difference exceeding the statistical uncertainty. Such relation
for masses may be rather general (model independent), as suggested
by comparison with the mass pairs (BW and pole) shown for $\pi N$
resonances in Listings of RPP~\cite{PDG}. The assumption is also
supported, if one solves Eq.(\ref{pol}) by decomposing in degrees
of $\Gamma_0/M_0\,$. In this way we obtain
\begin{equation}
	M_p-M_{BW}=-\frac14\,M_0\,(\Gamma_0/M_0)\,\Gamma_0^{\,\prime} ;
\label{pBWm}
\end{equation}
\begin{equation}
	\Gamma_p-\Gamma_{BW}=\frac18\,\Gamma_0\,(\Gamma_0/M_0)\,
	(3\Gamma_0^{\,\prime}-M_0\,\Gamma_0^{\,\prime\prime})\,.
\label{pBWg} \end{equation}
Here $\Gamma_0=\Gamma(M_0)\,$, while
$\Gamma_0^{\,\prime}\equiv d\,\Gamma(M_X)/d\,M_X\,,~~
\Gamma_0^{\,\prime\prime}\equiv d^2\,\Gamma(M_X)/(d\,M_X)^2$
taken at $M_X=M_0\,.$ Note that parametrically both $\Gamma_0^{\,
\prime}$ and $\Gamma_0^{\,\prime\prime}$ are  proportional to
$\Gamma_0\,.$ The three quantities $\Gamma_0/M_0,\,\Gamma_0^{\,
\prime},$ and $M_0\,\Gamma_0^{\,\prime\prime}$ are dimensionless
and, thus, have the same order of smallness, though may be very
different numerically.

If $M_0$ is not far from threshold of a decay channel, the
$\Gamma(M_X)$ is an increasing function near $M_0\,$, and
$\Gamma_0^{\,\prime}>0,$ providing $M_p<M_{BW}.$ Relation for the
widths is less definite. Due to the presence of both the first and
second derivatives, the sign of $\Gamma_p-\Gamma_{BW}$ may depend
on more details of the threshold behavior, which can be affected
by both the spin of the decaying resonance and by properties of
the final state. For the $\Lambda(1520)$, all three main
contributions (\ref{N}), (\ref{S}), and (\ref{L}) lead to
$\Gamma_p<\Gamma_{BW}\,,$ in accordance with our numerical
solution. It could be different for the case of $S$-wave
two-particle decays.

In summary, we have found the mass and width of the $\Lambda(1520)$
in near-forward electroproduction. The extracted Breit-Wigner
parameters of the resonance are shown to depend not only on
experimental data, but also on the way of treatment of the data.
The extracted width is more sensitive to the various treatments
than the mass. For the $\Lambda(1520)$, the non-resonance
background should be accurately studied and understood if one
intends to extract the mass and, especially, width with uncertainty
of order 1--2~MeV. 

Having the BW mass and width (\ref{MG}), we also give the first 
estimate (\ref{MpGp}) of the pole parameters for the $\Lambda(1520)\,.$ 
The pole values for both mass and width tend to be lower than the BW 
values.

In this Letter we have studied precision reachable for the basic 
parameters of the $\Lambda(1520)$. However, we could not, of course, 
advance here understanding of many questions related to this resonance.
Therefore, we urge its further investigations.   

\vspace{-5mm} \acknowledgments

The authors express their gratitude to R.~Arndt and to E.~Pasyuk
for useful discussions. One of the authors (Ya.~A.) highly
appreciates the hospitality and support extended to him by the
George Washington University and by the Jefferson Lab. This work
was supported in part by the U.~S.~Department of Energy under
Grants DE--FG02--99ER41110 and DE--FG02--03ER41231, by the Russian
State grant SS--3628.2008.2, and by Jefferson Science Associates
which operates the Thomas Jefferson National Accelerator Facility
under DOE contract DE-AC05-06OR23177.





\end{document}